# Nonlinear dynamic Process of Fluvial Process Based on Theories of Chaos and Dissipative Structure

Hao Lin

**Abstract**

The analysis of riverbed evolution is an important basis for decision-making on river improvement, flood control project construction, and water conservancy hub dispatching and application. From the perspective of river nonlinear dynamic system, this paper deeply understands the disorder and order in the evolution and development of river bed. Using the chaos theory, the phase diagram is legally used to identify the chaos of the river time series, and the fractal dimension is quantitatively calculated to calculate the characteristic quantities of chaos. Combined with the theory of dissipative structure, the orderliness in evolution and development is understood, and the internal mechanism of river self-organization is analyzed by the principle of minimum entropy generation. The river information entropy is introduced to identify the local scouring and silting pattern of the river section. The results show that the runoff and sediment transport in the three sections are chaotic, and the information entropy fluctuates with a certain range with time, and develops to a decreasing size, that is, the river section as a whole develops in an orderly direction. However, the chaos and disorder of the Sanmenxia section are the most prominent, which may be affected by the imbalance of sediment transport caused by the construction of the reservoir, and the change of beach trough scouring and silting here is significantly affected, and the situation has been improved with the development of bank protection projects.

**Introduction**

Under natural conditions (long-time-scale effects, such as climate change and crustal tectonic variations), rivers constantly undergo automatic adjustment and evolutionary development under the interaction of flow, sediment and riverbed. This evolution exhibits both disorder and intrinsic randomness, showing chaotic characteristics. With the increasing impact of human activities (such as the construction of large-scale water conservancy projects and sand mining), significant changes in the conditions of water and sediment supply and constraints are generally caused, which in turn disrupt the natural laws of riverbed evolution. However, the evolution process is affected by complex and variable factors. How to quantitatively identify the chaotic nature of river evolution and how to accurately assess and predict the impact of human activities on riverbed evolution remain the difficulties in the study of riverbed evolution. In this paper, the research on riverbed evolution is systematically summarized, and further combined with chaos theory, dissipative structure theory and self-organization theory to propose a riverbed evolution analysis model of qualitative identification, quantitative calculation and graphical analysis. The analysis model can be used to identify the chaotic and self-organization characteristics of rivers, as well as to analyze local riverbed anomalies. In view of the severe problems of river wandering and "secondary hanging river" in the Yellow River, the application of the systematic riverbed evolution analysis model is conducive to optimizing the operation and scheduling of reservoirs, analyzing and evaluating the impact of water conservancy projects, and has important engineering significance for comprehensive management.

# 1. Theory of Chaos

Previously, research on riverbed evolution predominantly relied on mathematical and physical models, which were limited to qualitative analysis or semi-quantitative calculations, resulting in low accuracy of the outcomes. With the introduction of chaos theory in the 1970s, deterministic systems with non-linear interactions garnered widespread attention. The seemingly random yet patterned phenomenon known as chaos has been identified in various aspects of river systems. A river system is a complex open system, characterized by its intrinsic randomness and sensitivity to initial conditions, which provide the fundamental conditions for the emergence of chaotic characteristics.

## 1.1 Overview of Chaos Theory in Rivers

Chaos theory, since its development in the 1970s, has found numerous applications in natural and social sciences. It has been demonstrated that river systems exhibit chaotic characteristics as non-linear dynamic systems. From the smallest scale of turbulent flow to the larger scales of water and sediment supply, and ultimately riverbed evolution, chaotic phenomena are omnipresent. Traditionally, chaos is used to describe a state of disorder and confusion. However, in the context of non-linear dynamic systems, it refers to a seemingly irregular motion that occurs under non-equilibrium conditions. This type of motion does not require any external random factors to exhibit similar random behavior (intrinsic randomness), yet it possesses a rich internal hierarchical structure of order. The chaos referred to in this paper is the chaos of dissipative systems, which are dynamic systems where energy varies over time and the evolution process is irreversible. Such dissipative systems with chaotic characteristics exhibit strange attractors.

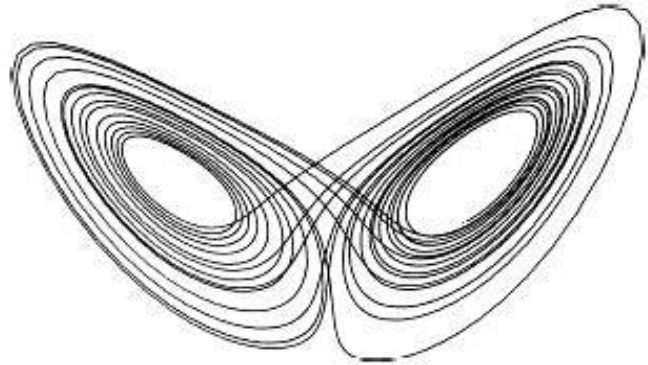

**Fig. 1.** Strange Attractors in the Chaotic System

Currently, research on the application of chaos theory in the field of water science primarily focuses on the analysis of chaotic characteristics in precipitation, runoff, floods, and sediment content. However, these studies have not addressed the chaotic characteristics of multivariate time series, whereas river systems are often influenced by numerous factors. The boundary conditions of the riverbed and the conditions of water and sediment supply together constitute the external conditions affecting riverbed evolution. The boundary conditions of the riverbed mainly include width-depth ratio, roughness, and slope, while the conditions of water and sediment supply mainly include runoff volume and sediment transport volume. In this study, the time series of runoff volume and sediment transport volume are selected for the analysis of chaotic characteristics.

## 1.2 Identification Methods for River Chaos Characteristics

To investigate the chaotic motion of non-linear dynamic systems, it is essential to first identify the chaos in the system. The methods can be broadly categorized into qualitative and quantitative approaches. Qualitative methods reveal the special spatial structures or frequency characteristics exhibited by chaotic motion trajectories in the time or frequency domain, such as chaotic strange attractors, to distinguish them from periodic, quasi-periodic, or random motions. These methods include phase diagram analysis, sub-frequency sampling, Poincaré sections, phase space reconstruction, power spectral analysis, and principal component analysis. Quantitative methods identify chaotic motion by calculating the characteristic quantities of chaotic strange attractors, such as fractal dimension, Lyapunov exponents, and measure entropy. In the identification of river chaos characteristics, phase diagram analysis and fractal dimension analysis are employed for qualitative and quantitative analyses, respectively. The existence of strange attractors is first identified through phase diagram analysis to qualitatively recognize chaotic characteristics, followed by the quantitative calculation of the characteristic quantities of the strange attractors using fractal dimension analysis to assess the degree of chaos in the river sections.

**(1) Phase Diagram Method**

The state quantities of a dynamic system at various moments constitute the phase space. The trajectories formed by the system's solutions over time in the phase space are referred to as phase trajectories, and the combination of phase trajectories and phase space forms the system's phase diagram. The phase diagram provides a qualitative description of the changes in the system's state over the study period and reflects the structure of strange attractors, thereby allowing for the qualitative analysis of the nature of the system's motion.

Taking the data from three hydrological stations (Longmen, Tongguan, and Sanmenxia) in the middle reaches of the Yellow River as an example, the monthly average runoff and sediment transport volumes from 2010 to 2020 were obtained based on the annual Yellow River Sediment Bulletins. By reconstructing the phase space, the two-dimensional phase diagrams were drawn as follows:

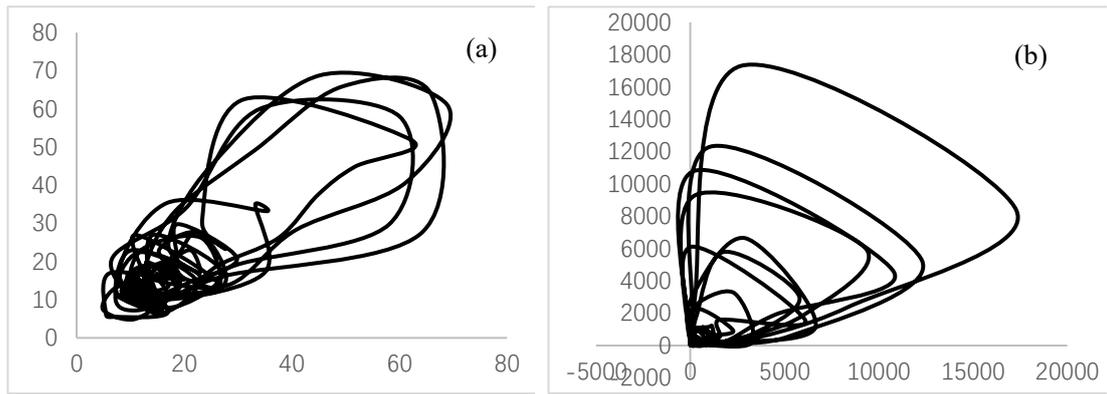

**Figure 2.** Two-dimensional phase diagram of runoff (a) and sediment transport (b) at the Longmen Hydrological Station

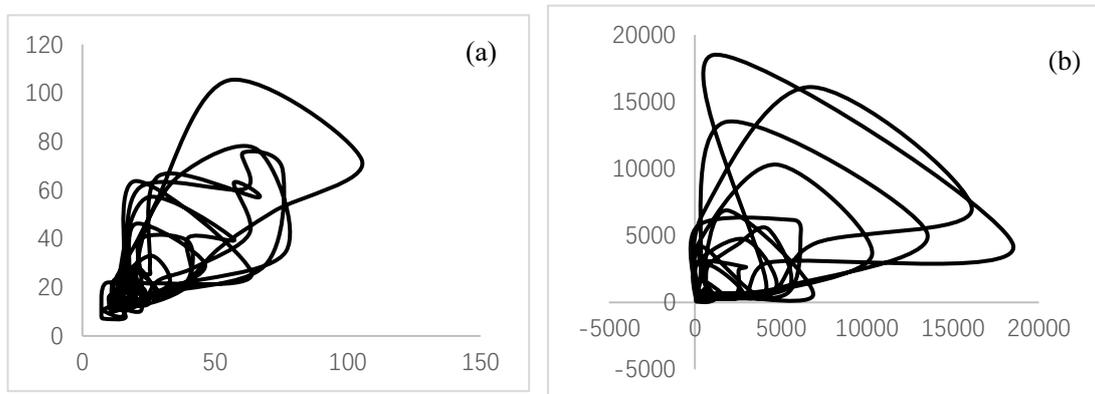

**Figure 3.** Two-dimensional phase diagram of runoff (a) and sediment transport (b) at the Tongguan Hydrological Station

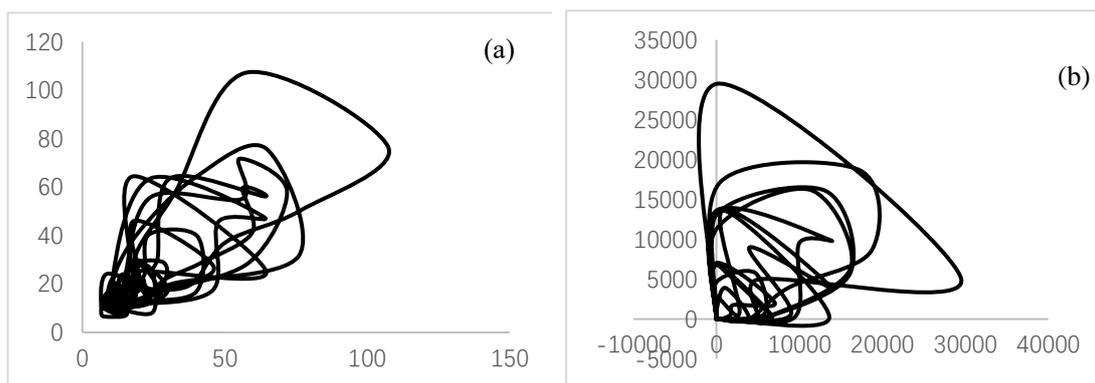

**Figure 4.** Two-dimensional phase diagram of runoff (a) and sediment transport (b) at the Sanmenxia Hydrological Station

From the two-dimensional phase diagrams of the monthly runoff and sediment transport time series at the three stations shown in Figures 2-4, it can be observed that the phase trajectories of the runoff and sediment transport at Longmen, Tongguan, and Sanmenxia stations all exhibit an attractor-centered attractor domain. The motion of each phase point continuously folds back and forth, approaching and moving away from the attractor domain. This is an important criterion for identifying chaotic characteristics. Therefore, it can be concluded that the time series of monthly runoff and sediment transport in this river section exhibit chaotic characteristics.

**(2) Fractal Dimension Method**

The chaotic characteristics can be described by the irregular trajectories of strange attractors. The geometric dimension of a strange attractor is fractional, hence the introduction of fractal dimension. Fractal dimension can quantitatively describe the geometric features of the attractor and the evolution of the trajectories on the attractor over time. Based on this, the degree of chaos of the attractor can be further subdivided. There are various representations of fractal dimension, such as Hausdorff dimension, similarity dimension, information dimension, and saturated correlation dimension. Among them, the saturated correlation dimension, which is based on the extraction of fractal dimension from experimental data, has been widely used as a method for identifying chaotic characteristics.

The saturated correlation dimension method first reconstructs an m-dimensional phase space based on the time series $\{x_i \mid i=1,2,...,N\}$, selecting a time delay $\tau$. The trajectory of the system in this m-dimensional phase space is then plotted using the measured time series. The reconstructed phase space can be expressed as

$$\begin{aligned} X_1 &= \{x_1, x_{1+\tau}, \cdots, x_{1+(m-1)\tau}\} \\ X_2 &= \{x_2, x_{2+\tau}, \cdots, x_{2+(m-1)\tau}\} \\ &\vdots \\ X_l &= \{x_l, x_{l+\tau}, \cdots, x_{l+(m-1)\tau}\} \end{aligned} \quad (1)$$

where $m$ is the embedding dimension of the phase space, $\tau$ is the delay time, $l$ is the total number of phase points, $l=N-(m-1)\tau$, and $X_1, X_2, \ldots, X_l$ are the phase space sequences.

In the m-dimensional phase space sequence $X_1, X_2, \ldots, X_l$, let $r_{ij}(m)$ be the absolute value of the difference between any two vectors (i.e., the Euclidean distance). Then,

$$r_{ij}(m) = \|X_i - X_j\| \quad (2)$$

Given a set of values $r_0$ within the range of the minimum and maximum values, by adjusting the size of $r_0$, a set of values for $\ln r_0$ and $\ln C(r)$ can be calculated. The correlation dimension $d_m$ can then be derived from the following equation:

$$d_m = \lim_{r \to 0} \frac{\ln C(r)}{\ln r_0} \quad (3)$$

where,

$$C(r) = \frac{1}{l(l-1)} \sum_{i,j=1, i \neq j}^{l} H(r_0 - r_{ij}) = \frac{1}{l(l-1)} \sum_{i,j=1, i \neq j}^{l} H(r_0 - \|X_i - X_j\|) \quad (4)$$

Here, $C(r)$ is the correlation integral, $l$ is the total number of phase points, and $H(x)$ is the Heaviside unit function, defined as:

$$H(x) = \begin{cases} 1, & x \geq 0 \\ 0, & x < 0 \end{cases} \quad (5)$$

Then, the relationship between $\ln r_0$ and $\ln C(r)$ is plotted for different embedding dimensions. It can be intuitively observed from the plot that if the curve of $\ln r_0$ versus $\ln C(r)$ contains a linear portion, the slope of this line is referred to as the correlation dimension $d_m$. In the presence of a strange attractor, as the embedding dimension $m$ increases, the correlation dimension *dm* also increases. When the correlation dimension reaches a certain value and becomes saturated, the saturated correlation dimension value *D* is obtained, which is the saturated correlation dimension of the time series. The larger the saturated correlation dimension, the stronger the chaotic characteristics; conversely, the smaller the saturated correlation dimension, the weaker the chaotic characteristics.

Taking the data from three hydrological stations (Longmen, Tongguan, and Sanmenxia) in the middle reaches of the Yellow River as an example, the monthly average runoff and sediment transport volumes from 2010 to 2020 were obtained based on the annual Yellow River Sediment Bulletins. The saturated correlation dimension was calculated using Equations (1) to (5), and the relationship plots are shown in Figures 5-7, which are used to quantitatively assess the degree of chaos at each station in the study river section.

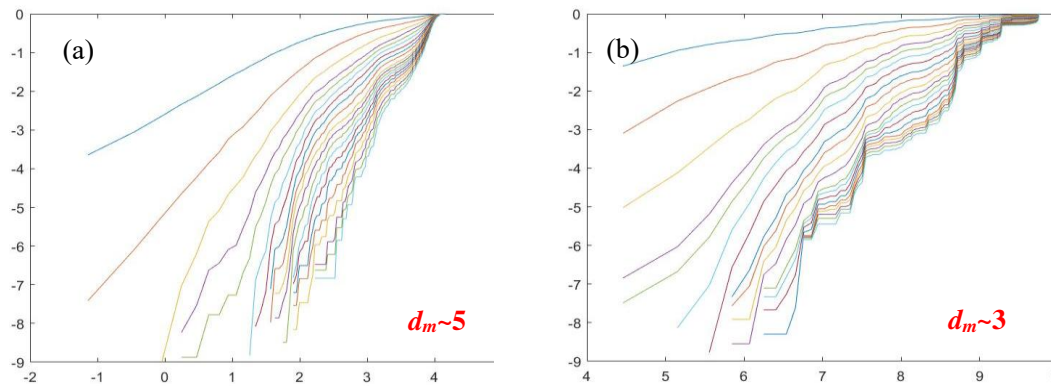

**Figure 5.** The relationship between the runoff (a) and the sediment load (b) of the Longmen hydrological station

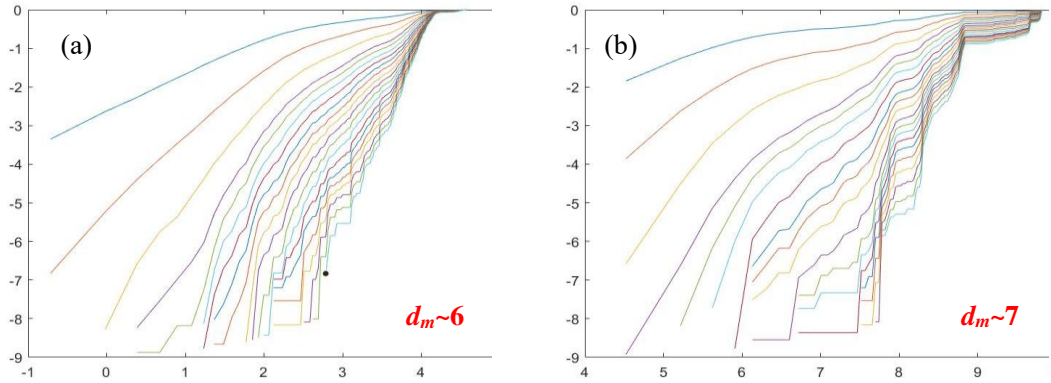

**Figure 6.** The relationship between the runoff (a) and the sediment load (b) of the Tongguan hydrological station

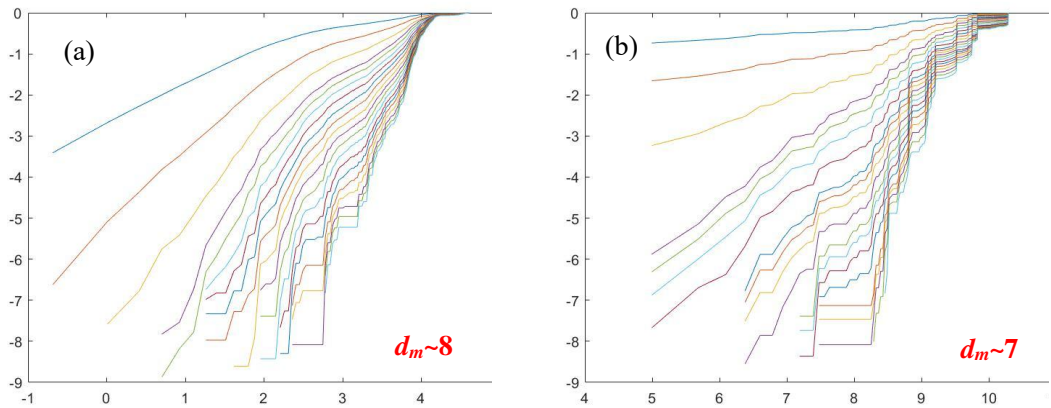

**Figure 7.** The relationship between the runoff (a) and the sediment load (b) of the Sanmenxia hydrological station

From the relationship plots in Figures 5-7, it can be seen that there is a linear portion in the ln$r$0-ln$C(r)$ relationship for different embedding dimensions. Therefore, the monthly runoff and sediment transport time series in this section of the Yellow River exhibit chaotic characteristics. For the Longmen station, the correlation dimension of the monthly runoff and sediment transport sequences no longer increases with the embedding dimension after $m≈3$ and $m≈5$, so the saturated correlation dimension is taken as 4. For the Tongguan station, the correlation dimension of the monthly runoff and sediment transport sequences no longer increases with the embedding dimension after $m \approx 6$ and $m \approx 7$, so the saturated correlation dimension is taken as 6. For the Sanmenxia station, the correlation dimension of the monthly runoff and sediment transport sequences no longer increases with the embedding dimension after $m \approx 7$ and $m \approx 8$, so the saturated correlation dimension is taken as 8. Thus, it can be preliminarily concluded that the river section from Tongguan to Sanmenxia has the highest degree of chaos, while the section from Longmen to Tongguan has the lowest degree of chaos. The high degree of chaos in the Sanmenxia section reflects the relatively intense riverbed evolution there, which may be related to the water and sediment regulation by the Sanmenxia Reservoir and the long-term sediment accumulation in front of the reservoir. In contrast, the upstream section is relatively stable with a gentle river trend.

**2. Dissipative Structure Theory**

Chaos theory reveals the state of disorder within the system, but at the same time, the river system also exhibits a certain order, that is, the dissipative structure. The dissipative structure and chaotic state alternate in the evolution of the riverbed, and the river adjusts its own shape (such as increasing the river width, decreasing the slope drop, etc.) through self-organization under the external constraints, so as to achieve the relative equilibrium state of the river, and at the same time, different river shapes and development laws may be formed. In addition, in terms of spatial pattern, the local erosion and siltation changes of rivers can be quantitatively calculated by information entropy, and then analyzed by combining remote sensing images.

## 2.2 Dissipative Structure Theory

The theory of dissipative structures was first proposed by Prigogine in 1969, referring to the ordered structures that emerge in a system when a certain parameter undergoes fluctuations and develops through non-equilibrium phase transitions. Rivers are open systems that exchange energy and matter with their surroundings while also interacting through information entropy. The external environment influences the direction of river development through positive or negative entropy flows. Negative entropy flows promote the development of rivers towards order, while positive entropy flows drive rivers towards disorder. When an open system is far from equilibrium, it has multiple developmental directions, which in thermodynamic terms is explained as bifurcation or branching phenomena. Through bifurcation and mutation, two different states may be reached: one is a dissipative structure, and the other is a chaotic state. Dissipative structures and chaotic states alternate during the transformation of river types, leading to the formation of different river patterns. The theory of dissipative structures also attempts to understand the internal self-organization mechanisms and patterns of systems to some extent.

## 2.2 River Self-Organization Theory

The theory of dissipative structures was first proposed by Prigogine in 1969, referring to the ordered structures that emerge in a system when a certain parameter undergoes fluctuations and develops through non-equilibrium phase transitions. Rivers are open systems that exchange energy and matter with their surroundings while also interacting through information entropy. The external environment influences the direction of river development through positive or negative entropy flows. Negative entropy flows promote the development of rivers towards order, while positive entropy flows drive rivers towards disorder. When an open system is far from equilibrium, it has multiple developmental directions, which in thermodynamic terms is explained as bifurcation or branching phenomena. Through bifurcation and mutation, two different states may be reached: one is a dissipative structure, and the other is a chaotic state. Dissipative structures and chaotic states alternate during the transformation of river types, leading to the formation of different river patterns. The theory of dissipative structures also attempts to understand the internal self-organization mechanisms and patterns of systems to some extent.

## 2.3 River Information Entropy Theory

Since Clausius first introduced the concept of entropy in 1854, it has been widely applied in various fields. Entropy is a fundamental concept in thermodynamics, used to represent a state function of a system. In 1948, Shannon first introduced the concept of entropy into information theory, calling it information entropy, to represent the degree of order in a system. Information is a measure of the orderliness of a system, while information entropy is a measure of the disorderliness of a system. In river systems, information entropy has been widely used to evaluate the disorderliness of river patterns and the wandering nature of riverbeds. Research has shown that the greater the information entropy, the more turbulent the river pattern and the more scattered the river. Previous studies have mostly focused on two-dimensional space in the time dimension, with few considering the three-dimensional spatiotemporal scale of river sections. A three-dimensional spatiotemporal analysis of riverbed evolution is conducive to identifying local scouring and silting changes in river sections and studying the patterns of evolutionary development.

The main factors affecting riverbed evolution can be divided into water and sediment supply conditions and riverbed boundary conditions. Due to the limitations of available data, this study selects runoff and sediment transport volumes for indicator calculation and analysis, without effectively characterizing the riverbed boundary conditions.

For an uncertain system, if a random variable $X$ represents its state characteristics, and for a discrete random variable, let the values of $X=\{x_1, x_2, \cdots, x_n\}(n \geq 2)$, with corresponding probabilities $P=\{p_1, p_2, \cdots, p_n\}$ ($0 \leq p_i \leq 1, i=1,2,\cdots,n$), and $\sum_{i=1}^{n} p_i = 1$ the information entropy can be calculated as follows:

(1) Construct the initial matrix for $n$ years and $m$ influencing factors $X_{ij} = [x_{ij}](i=1,2,\cdots,n; j=1,2,\cdots,m)$.

(2) Normalize the influencing factors. Since the influencing factors represent different objects with different dimensions and orders of magnitude, the original data need to be normalized. Let the influencing factors of the river section be denoted as $Xij$. The normalized value $Xij$ is calculated as:

$$X_{ij}^* \begin{cases} \dfrac{x_{ij} - x_{j,\min}}{x_{j,\max} - x_{j,\min}} \\ \dfrac{x_{j,\max} - x_{ij}}{x_{j,\max} - x_{j,\min}} \end{cases} \quad (6)$$

where $xij*$ is the normalized value, $xij$ is the actual measured value of the $j$-th influencing factor in the $i$-th year, $xj,\max$ and $xj,\min$ are the maximum and minimum values of the $j$-th influencing factor over all years, respectively.

(3) Calculate the information entropy of each influencing factor using the formula:

$$S_j = -\frac{1}{\ln n} \sum_{i=1}^{n} \frac{x_{ij}^*}{x_j} \ln(\frac{x_{ij}^*}{x_j}) \qquad (7)$$

where $S_j$ is the information entropy of the $j$-th influencing factor, $x_{ij^*}$ is the normalized value of the $j$-th influencing factor in the $i$-th year, and $x_j$ is the sum of the $j$-th influencing factor over all years $x_j = \sum_{i=1}^{n} x_{ij}^* \ (i=1,2,\cdots,n; j=1,2,\cdots,m)$.

(4) Calculate the information entropy weight of each influencing factor using the formula:

$$\omega_j = \frac{1-S_j}{m - \sum_{j=1}^{m} S_j} \qquad (8)$$

(5) Combine the weights and values of the influencing factors to calculate the comprehensive evaluation entropy for each time point using the formula:

$$E_i = \sum_{j=1}^{m} \omega_j x_{ij} \qquad (9)$$

Taking the data from three hydrological stations (Longmen, Tongguan, and Sanmenxia) in the middle reaches of the Yellow River as an example, the monthly average runoff and sediment transport volumes from 2010 to 2020 were obtained based on the annual Yellow River Sediment Bulletins. The information entropy for each station over the 10 years was calculated using Equations (6) to (9), and the results are shown in Figure 8. The comprehensive evaluation index values (with runoff and sediment transport volumes as indicators) for each station were also calculated and plotted in Figure 9.

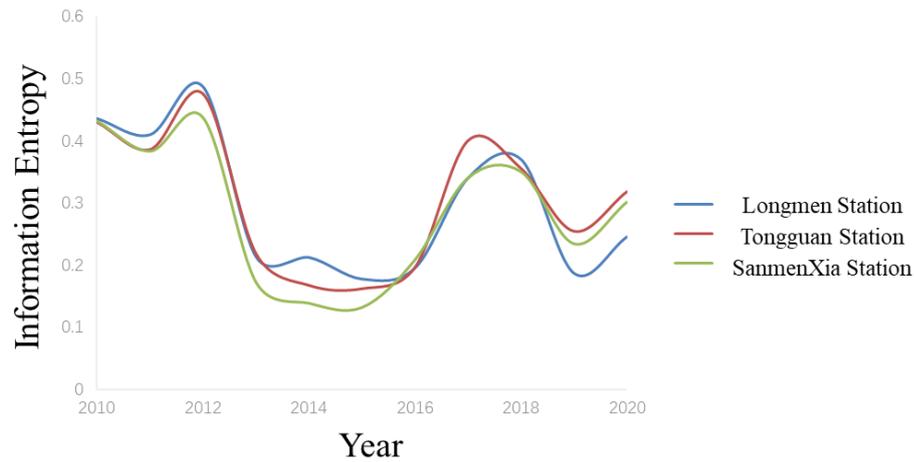

**Figure 8.** The relationship between information entropy and time change of Longmen, Tongguan, and Sanmenxia reaches

From Figure 8, it can be seen that the information entropy trends of the three river sections are similar, with overall fluctuations over time and values below 0.6. The information entropy at the three stations reached a relatively high value of approximately 0.51 around 2012, which may have been caused by climatic conditions of runoff that year leading to unbalanced sediment transport, or it may have been related to human activities, resulting in short-term unstable river trends. Between 2013 and 2015, there were relatively low values of about 0.2, which may have been due to the construction of river regulation projects during that period, which played a role in stabilizing the river trends. More importantly, the river could adjust itself through its self-organization capabilities to adapt to the changing boundary conditions. Subsequently, between 2016 and 2018, the information entropy reached a secondary high point of about 0.4, which was smaller than the previous peak (about 0.51). This indicates that under the new boundary conditions, the river reached a new equilibrium state through its own adjustment, with a lower degree of disorder and a more stable river trend. Information entropy is a measure of the disorderliness of a system. Based on the above analysis, it can be concluded that the three river sections of the Yellow River overall showed an orderly trend over time, with occasional high disorderliness in individual years. However, in the short term, the construction of bank protection projects could effectively control the disorderliness (scattering) of the river sections. In the long term, the river would adjust itself through self-organization to adapt to the new boundary conditions, thereby "restoring" stability and reaching a new equilibrium state. Therefore, the information entropy trends of the three river sections from 2010 to 2020 were similar, with overall fluctuations within an elastic range, indicating that the river trends in the study section of the middle reaches of the Yellow River were generally stable.

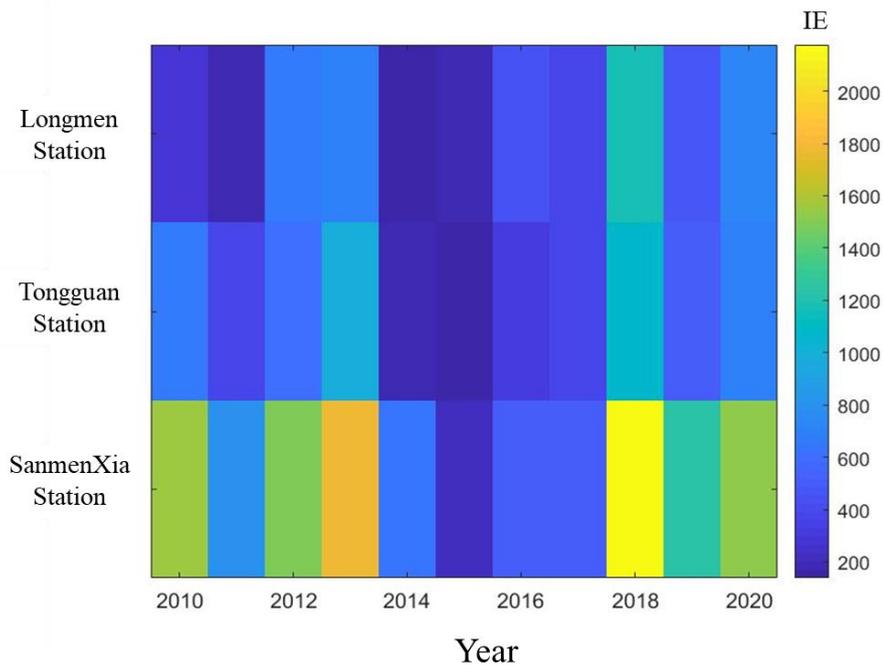

**Figure 9.** The relationship between information entropy (IE) and spatial variation of Longmen, Tongguan, and Sanmenxia reaches

Figure 9 shows the value of information entropy (IE) in various locations. It can be seen that there are local high-entropy river sections (Sanmenxia section), revealing local uncertainty and disorder in the riverbed, reflecting the wandering and scattered characteristics of the river in that section. The possible cause may be the construction of the Sanmenxia Reservoir, which has made the sediment transport imbalance in that section particularly prominent, leading to local river trend instability. Moreover, the river section is located in the middle and lower reaches of the Yellow River, where the wandering characteristics are significant, and the river is scattered. From the actual data and remote sensing images, it was observed that the local changes in the shoal and channel scouring and silting in the Longmen section were significant from 2010 to 2020, but the information entropy changes were small. This may be due to the fact that this study only considered runoff and sediment transport volumes and did not calculate the riverbed boundary conditions, which has certain limitations.

## 3. Conclusions

(1) River systems are nonlinear dynamic systems. By applying chaos theory to reconstruct the phase space of river time series, chaotic characteristics of the monthly runoff were qualitatively identified through the observation of strange attractors using phase diagram analysis. Subsequently, the degree of chaos in the river sections was quantitatively assessed by calculating the characteristic quantities of the strange attractors using fractal dimension analysis. The case study calculations revealed that the sections from Longmen to Tongguan to Sanmenxia in the middle reaches of the Yellow River all exhibit chaotic characteristics, with the section from Tongguan to Sanmenxia showing the highest degree of chaos and the section from Longmen to Tongguan exhibiting the lowest degree of chaos.

(2) The theory of dissipative structures was employed to further comprehend the ordered nature within nonlinear dynamic systems. By integrating the principle of minimum entropy production, the patterns and regulatory functions of river self-organization were explored. Rivers adjust their forms and structures to adapt to external constraints, thereby minimizing entropy production.

(3) The concept of river information entropy was introduced to conduct spatiotemporal analyses of riverbed evolution. Spatially, local changes in scouring and silting, as well as shoal and channel anomalies, were quantitatively identified. The Sanmenxia section was found to have higher information entropy compared to the Tongguan and Longmen stations, indicating a greater degree of disorder and potentially more intense scouring and silting. This finding was corroborated by remote sensing images, which revealed significant changes in shoal and channel scouring and silting approximately 10 km upstream of the Sanmenxia station. Temporally, the information entropy trends of the three river sections were similar, with fluctuations within a certain range from 2010 to 2020. Overall, the entropy values exhibited a decreasing trend, suggesting that the Xiaobei mainstream section of the Yellow River is evolving towards a more ordered state.